\documentclass[12pt]{article}
\usepackage{graphicx}
\usepackage{mathtools}
\usepackage{cite}

\newcommand\pubnumber{TTK-21-59}
\newcommand\pubdate{\today}

\def\Title#1{\begin{center} {\Large #1 } \end{center}}
\def\Author#1{\begin{center}{ \sc #1} \end{center}}
\def\Address#1{\begin{center}{ \it #1} \end{center}}

\newcommand\pubblock{\rightline{\begin{tabular}{l} \pubnumber\\
         \pubdate  \end{tabular}}}
\newenvironment{Abstract}{\begin{quotation}  }{\end{quotation}}
\newenvironment{Presented}{\begin{quotation} \begin{center} 
             PRESENTED AT\end{center}\bigskip 
      \begin{center}\begin{large}}{\end{large}\end{center} \end{quotation}}
\def\Acknowledgements{\bigskip  \bigskip \begin{center} \begin{large}
             \bf ACKNOWLEDGEMENTS \end{large}\end{center}}

\begin{document}
\begin{titlepage}
\pubblock

\vfill
\Title{Top-pair events with B-hadrons at the LHC}
\vfill
\Author{Terry Generet}
\Address{Institut f\"ur Theoretische Teilchenphysik und Kosmologie,\\RWTH Aachen University, D-52056 Aachen, Germany}
\vfill
\begin{Abstract}
In these proceedings, we summarise the results of the recent calculation of the NNLO QCD corrections for the production of a top-quark pair in association with a bottom-flavoured hadron. The results consist of differential distributions of observables involving the identified hadron.
\end{Abstract}
\vfill
\begin{Presented}
$14^\mathrm{th}$ International Workshop on Top Quark Physics\\
(videoconference), 13--17 September, 2021
\end{Presented}
\vfill
\end{titlepage}
\def\thefootnote{\fnsymbol{footnote}}
\setcounter{footnote}{0}

\section{Introduction}

Since the discovery of the top quark, there has been much interest in measuring its mass with ever increasing precision. The top-quark mass is one of the fundamental parameters of the Standard Model, with ties to several important aspects, such as electroweak symmetry breaking and the stability of the vacuum. As a result, many different approaches to measuring the top-quark mass have been proposed and studied over the years. These are typically centred around observables involving the top quark's decay products, e.g.\ a $b$-jet.

However, the experimental precision of such observables is often limited by jet energy scale uncertainties (see e.g.\ ref.\ \cite{Aad:2015nba}). If one replaces the $b$-jet with a bottom-flavoured hadron, then one can essentially eliminate this source of uncertainty (see e.g.\ ref.\ \cite{Khachatryan:2016pek}). For this reason, the decay of a top quark to an identified hadron has been studied extensively in the past though next-to-leading order (NLO) in QCD\cite{Kharchilava:1999yj,Biswas:2010sa,Agashe:2012bn,Agashe:2016bok}.

Such calculations can be performed by using standard tools to describe the production of partons, followed by the fragmentation of those partons to the desired hadron as described by fragmentation functions (FFs)\cite{Berman:1971xz}. In particular, the production of heavy-flavoured hadrons, including $B$-hadrons, can be described using the perturbative fragmentation function formalism\cite{Mele:1990cw}. In these proceedings, we summarise the main results of ref.\ \cite{Czakon:2021ohs}, which presents the first NNLO QCD calculation of top-quark pair production and decay in association with a $B$-hadron. A more elaborate discussion of the concepts related to fragmentation is provided in the original article.

The structure of these proceedings is as follows. In section \ref{sec:2}, the main results are highlighted and briefly discussed. We conclude in section \ref{sec:3}. On top of several additional results, all relevant input can be found in the original article\cite{Czakon:2021ohs}, including how the FFs were constructed and all the parameters of the numerical computation.

\section{Results}\label{sec:2}
The results presented here are the differential distributions of three observables: the invariant mass of the $B$-hadron with the charged lepton $\ell$ ($m(B\ell)$), the energy of the $B$-hadron in the lab frame ($E(B)$) and the ratio of the transverse momenta of the $B$-hadron and the jet that contains it ($p_T(B)/p_T(j_B)$). In order to obtain these results, the \textsc{Stripper} framework -- a \textsc{c++} implementation of the four-dimensional formulation of the sector-improved residue subtraction scheme\cite{Czakon:2010td,Czakon:2011ve,Czakon:2014oma,Czakon:2019tmo} -- has been extended to be able to perform calculations involving fragmentation.

The only phase-space cuts applied are
\begin{equation}
p_T(B) > 10\text{ GeV}\:,\;\;\;\; |\eta(B)| < 2.4\:.
\end{equation}
For $p_T(B)/p_T(j_B)$, a jet definition must be specified. For this purpose, the anti-$k_T$ algorithm with $R = 0.8$ was used.

\begin{figure}[t]
	\centering
	\includegraphics[width=0.49\textwidth]{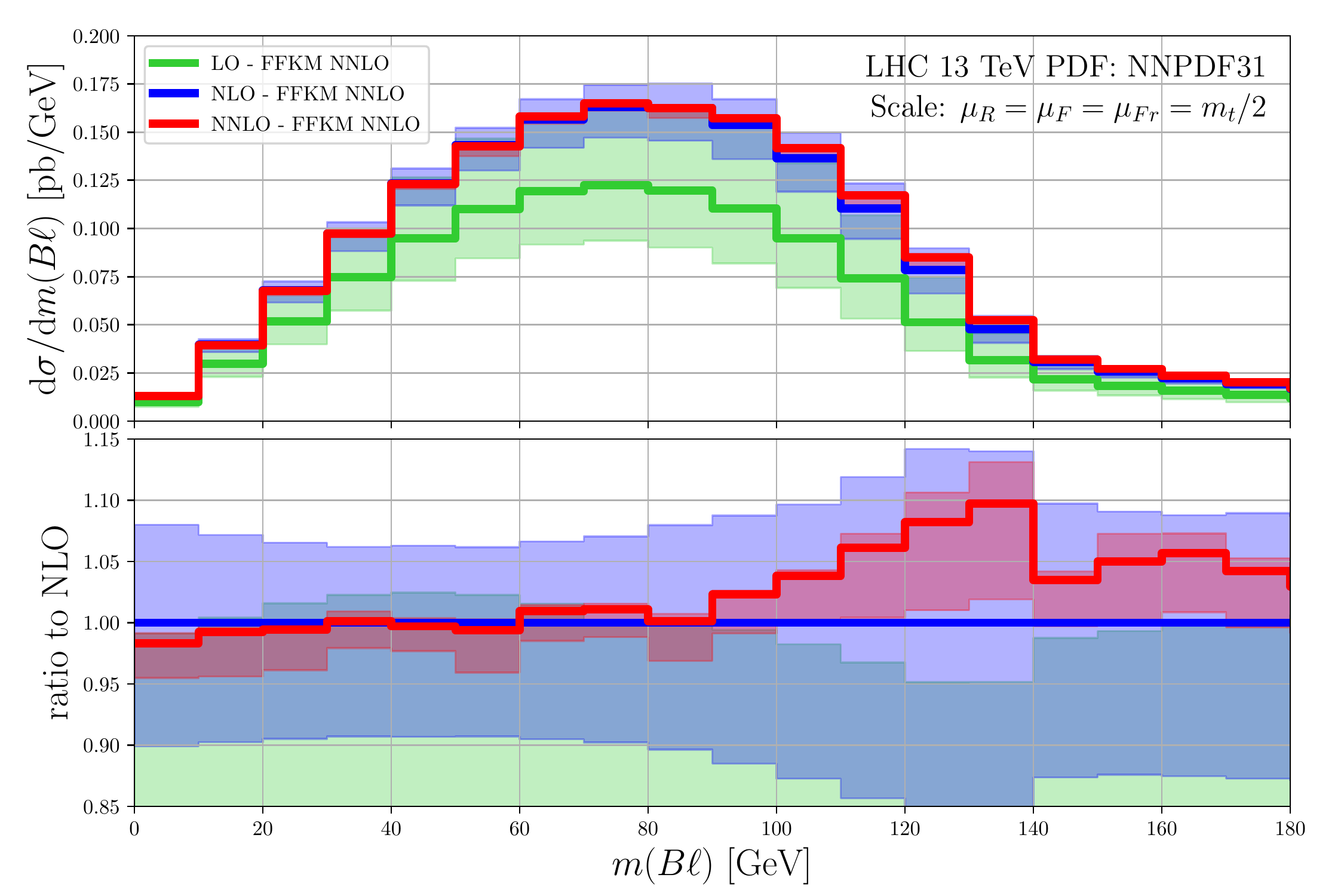}
	\includegraphics[width=0.49\textwidth]{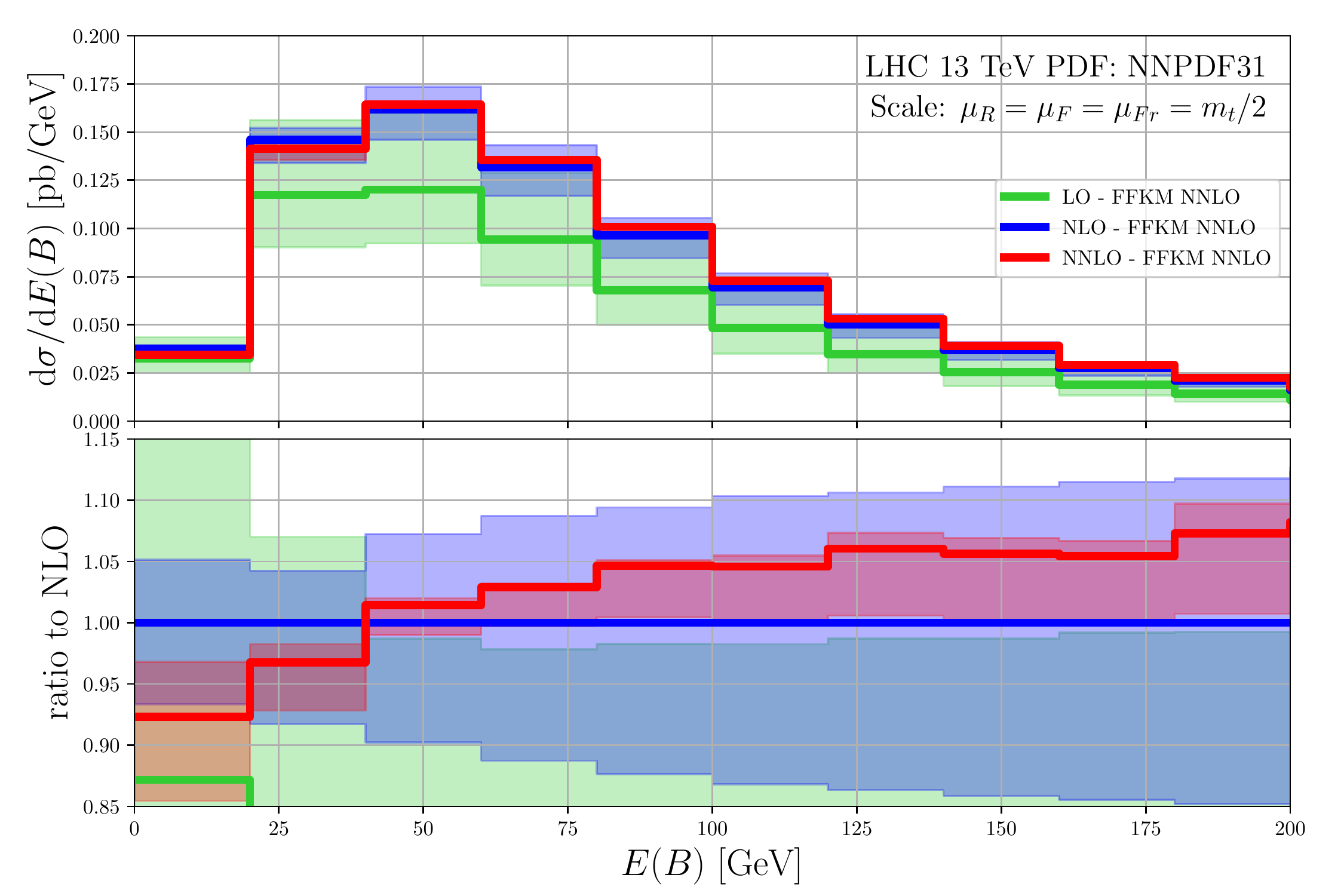}
	\caption{Absolute differential top-quark pair production and decay cross section as a function of the invariant mass $m(B\ell)$ (left) and the $B$-hadron energy $E(B)$ (right).}
	\label{fig:ttbar-order}
	\vskip 8mm
	\includegraphics[width=0.49\textwidth]{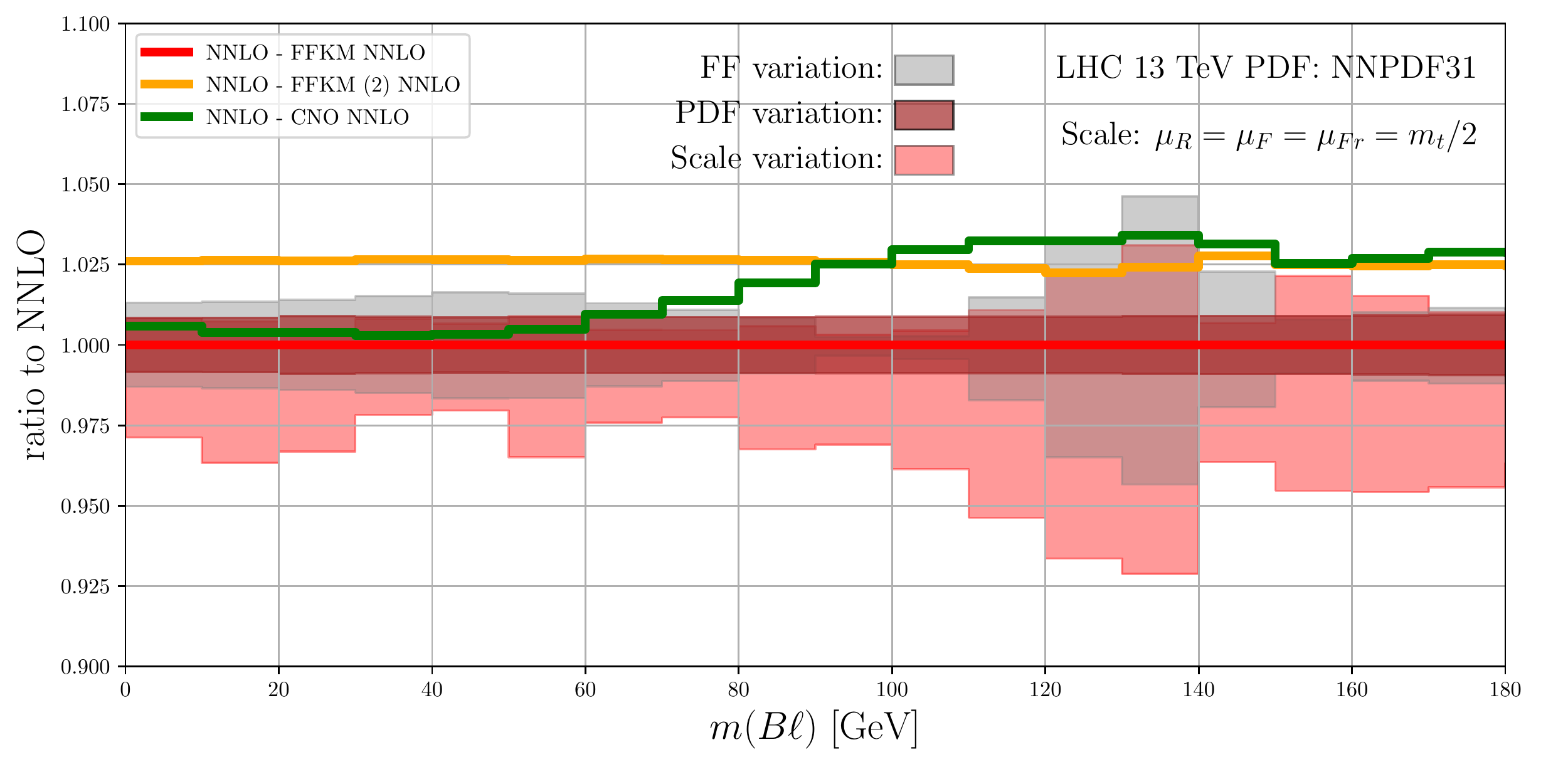}
	\includegraphics[width=0.49\textwidth]{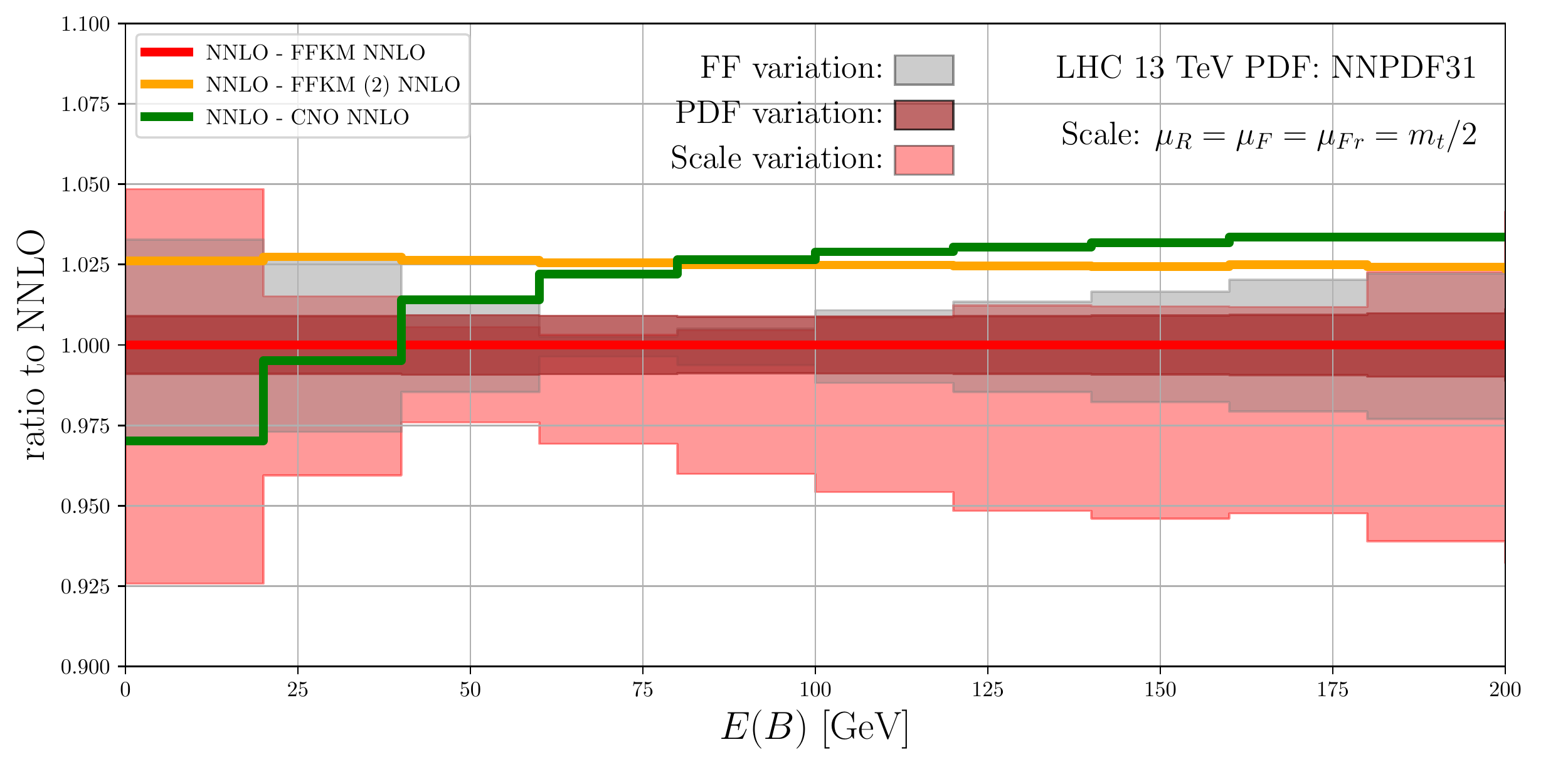}
	\caption{As in fig.~\ref{fig:ttbar-order} but showing the fragmentation and PDF variations of the default FF set. Also shown are the central predictions for two alternative FF sets.}
	\label{fig:ttbar-fragmentation}
\end{figure}

Fig.\ \ref{fig:ttbar-order} shows a comparison of the LO, NLO and NNLO spectra of $m(B\ell)$ and $E(B)$. In both cases, the NNLO corrections are considerably smaller than the NLO ones. In fact, the NNLO corrections lie completely within the NLO scale bands. The scale uncertainties are also significantly reduced at NNLO compared to NLO, leading to a residual scale uncertainty of a few percent in each bin. Fig.\ \ref{fig:ttbar-fragmentation} shows a comparison of different sources of uncertainty for the same two observables: the NNLO scale uncertainty, the uncertainty due to the FF and the PDF error. Also shown are the central curves of the NNLO predictions using two alternative FF sets. Scale uncertainties are still the dominant source of uncertainty at NNLO, followed by the FF errors. The PDF errors are small and almost perfectly flat.

\begin{figure}[t]
	\centering
	\includegraphics[width=0.5\textwidth]{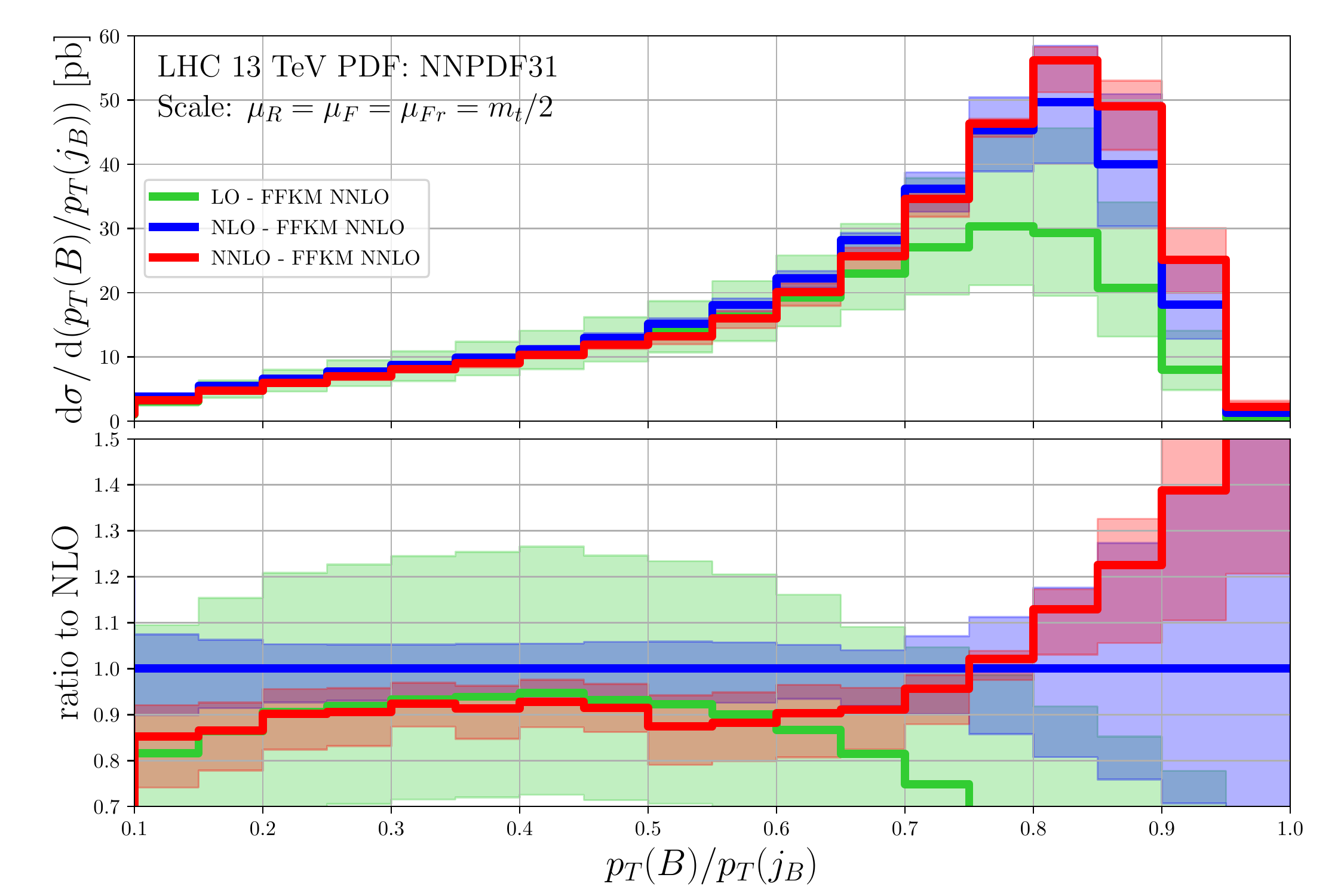}%
	\includegraphics[width=0.5\textwidth]{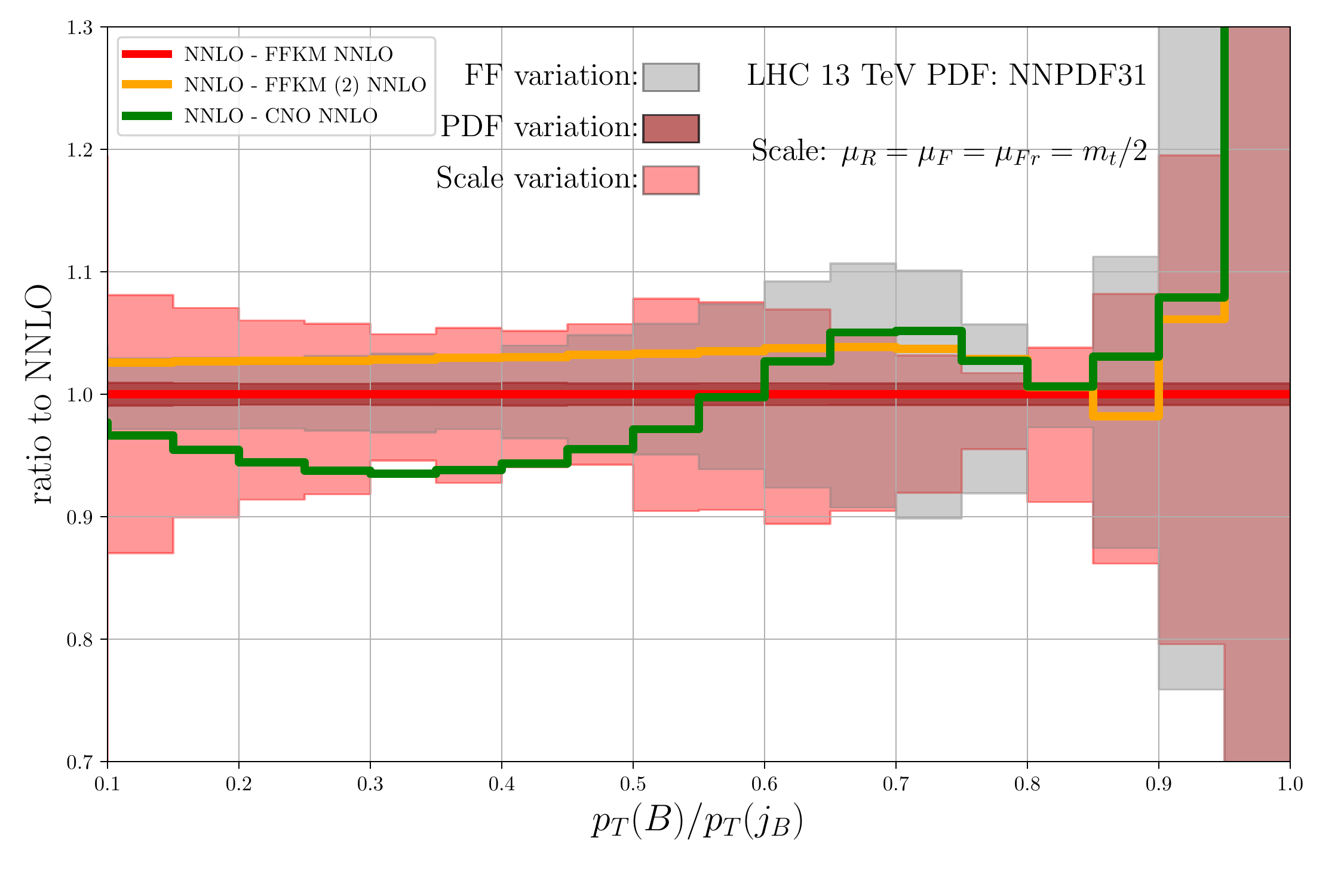}
	\caption{Absolute differential cross section as a function of the transverse momentum ratio $p_T(B)/p_T(j_B)$ in top-quark pair production and decay as in figs.\ \ref{fig:ttbar-order} and \ref{fig:ttbar-fragmentation}.}
	\label{fig:prod_jetratio}
\end{figure}

In analogy to figs.\ \ref{fig:ttbar-order} and \ref{fig:ttbar-fragmentation}, fig.\ \ref{fig:prod_jetratio} shows a comparison of the LO, NLO and NNLO spectrum of $p_T(B)/p_T(j_B)$, as well as a comparison of the different sources of uncertainty. This observable is of particular interest, because it is very close to the definition of the fragmentation function and therefore an excellent candidate for extractions of fragmentation functions from LHC data. However, one concern is that the shape of the spectrum might depend on the PDF set used, thus yielding different extracted FF sets depending on which PDF set was used in the analysis. Fig.\ \ref{fig:prod_jetratio} demonstrates that the PDF-dependence of this observable is not only tiny, but once again almost perfectly flat and that a PDF-insensitive extraction of FFs at the LHC is thus plausible.

\section{Conclusion}\label{sec:3}
In these proceedings, we have reported on the recent calculation of the NNLO QCD corrections to top-quark pair production in association with a $B$-hadron at the LHC. We have focused on the main results, which are differential distributions of observables involving the $B$-hadron for the full top-quark pair production and decay process. We found that the NNLO corrections are small and significantly reduce the scale uncertainties. It has also been demonstrated that extracting fragmentation functions from LHC data in a PDF-insensitive way may be possible. This work is a first application of a newly implemented numerical framework for computing NNLO QCD cross sections involving fragmentation, which is completely general and can thus be readily applied to a large variety of processes in the future.

\Acknowledgements
The work of T.G. was supported by the Deutsche Forschungsgemeinschaft (DFG) under grant 400140256 - GRK 2497: The physics of the heaviest particles at the Large Hadron Collider.


\begin{thebibliography}{99}
	
	\bibitem{Aad:2015nba}
	G.~Aad \textit{et al.} [ATLAS],
	Eur. Phys. J. C \textbf{75}, no.7, 330 (2015)
	[arXiv:1503.05427 [hep-ex]].
	
	\bibitem{Khachatryan:2016pek}
	V.~Khachatryan \textit{et al.} [CMS],
	JHEP \textbf{12}, 123 (2016)
	[arXiv:1608.03560 [hep-ex]].
	
	\bibitem{Kharchilava:1999yj}
	A.~Kharchilava,
	Phys. Lett. B \textbf{476}, 73-78 (2000)
	[arXiv:hep-ph/9912320 [hep-ph]].
	
	\bibitem{Biswas:2010sa}
	S.~Biswas, K.~Melnikov and M.~Schulze,
	JHEP \textbf{08}, 048 (2010)
	[arXiv:1006.0910 [hep-ph]].
	
	\bibitem{Agashe:2012bn}
	K.~Agashe, R.~Franceschini and D.~Kim,
	Phys. Rev. D \textbf{88}, no.5, 057701 (2013)
	[arXiv:1209.0772 [hep-ph]].
	
	\bibitem{Agashe:2016bok}
	K.~Agashe, R.~Franceschini, D.~Kim and M.~Schulze,
	Eur. Phys. J. C \textbf{76}, no.11, 636 (2016)
	[arXiv:1603.03445 [hep-ph]].
	
	\bibitem{Berman:1971xz}
	S.~M.~Berman, J.~D.~Bjorken and J.~B.~Kogut,
	Phys. Rev. D \textbf{4}, 3388 (1971)
	
	\bibitem{Mele:1990cw}
	B.~Mele and P.~Nason,
	Nucl. Phys. B \textbf{361}, 626-644 (1991)
	[erratum: Nucl. Phys. B \textbf{921}, 841-842 (2017)]
	
	\bibitem{Czakon:2021ohs}
	M.~Czakon, T.~Generet, A.~Mitov and R.~Poncelet,
	JHEP \textbf{10}, 216 (2021)
	doi:10.1007/JHEP10(2021)216
	[arXiv:2102.08267 [hep-ph]].
	
	\bibitem{Czakon:2010td}
	M.~Czakon,
	Phys. Lett. B \textbf{693}, 259-268 (2010)
	[arXiv:1005.0274 [hep-ph]].
	
	\bibitem{Czakon:2011ve}
	M.~Czakon,
	Nucl. Phys. B \textbf{849}, 250-295 (2011)
	[arXiv:1101.0642 [hep-ph]].
	
	\bibitem{Czakon:2014oma}
	M.~Czakon and D.~Heymes,
	Nucl. Phys. B \textbf{890}, 152-227 (2014)
	[arXiv:1408.2500 [hep-ph]].
	
	\bibitem{Czakon:2019tmo}
	M.~Czakon, A.~van Hameren, A.~Mitov and R.~Poncelet,
	JHEP \textbf{10}, 262 (2019)
	[arXiv:1907.12911 [hep-ph]].
	
\end{thebibliography}
\end{document}